# Polynomial Fourier Domain as a Domain of Signal Sparsity

Srdjan Stanković, *Senior Member, IEEE,* Irena Orović, *Member, IEEE,* Ljubiša Stanković, *Fellow, IEEE*,

*Abstract*— A compressive sensing (CS) reconstruction method for polynomial phase signals is proposed in this paper. It relies on the Polynomial Fourier transform, which is used to establish a relationship between the observation and sparsity domain. Polynomial phase signals are not sparse in commonly used domains such as Fourier or wavelet domain. Therefore, for polynomial phase signals standard CS algorithms applied in these transformation domains cannot provide satisfactory results. In that sense, the Polynomial Fourier transform is used to ensure sparsity. The proposed approach is generalized using time-frequency representations obtained by the Local Polynomial Fourier transform (LPFT). In particular, the first-order LPFT can produce linear time-frequency representation for chirps. It provides revealing signal local behavior, which leads to sparse representation. The theory is illustrated on examples.

## I. INTRODUCTION

Compressive sensing (CS) algorithms are developed to deal with an incomplete set of randomly selected/acquired signal samples [1]-[4]. The number of available samples is usually far below that required by the Nyquist sampling theorem. The corresponding hardware prototypes performing the sub-Nyquist sampling of analog inputs have recently been designed [5],[6]. In order to successfully reconstruct a full-length signal (i.e., original signal samples) from its small set of observations (CS samples), the signal should have a dense representation in the sensing domain and a sparse representation in a certain transform domain (i.e., sparsity domain). For instance, the Fourier transform domain can be considered as a sparsity domain for the case of signals that consist of $K$ sinusoids, where $K \ll M$ ($M$ represents the signal length). In this case, some of the common CS algorithms such as greedy, basis pursuit or recently proposed L-estimate CS algorithms can be employed to produce an exact reconstruction of original signal [7]-[21]. A significant contribution to the CS theory, especially in terms of reducing the number of necessary measurements, has been provided by the model-based CS approaches [15],[16]. These approaches explore the inter-dependency structure of large coefficients called the structured sparsity model. An example is tree-sparsity of wavelet coefficients in natural images [17]-[19], having largest coefficients clustered along the tree structure.

In real applications we usually deal with signals whose phase can be modeled by a certain nonlinear function over a limited time window. Such signals are not sparse in the Fourier, Wavelet or Fractional Fourier domain, but rather time-frequency or the Fractional Fourier domain [22]-[29]. However, if the time-frequency domain is used, then an appropriate distribution should be employed to provide sparse signal representation. For instance, if the short-time Fourier transform is used, then the chirp signal does not have a sparse representation and cannot be exactly reconstructed from random observations [30]-[34]. Furthermore, if the Wigner distribution is considered, only the monocomponent signal can be treated (to avoid the influence of cross-terms). Also, it is necessary to calculate the autocorrelation function [35], [36], which means that the number of available observations must be greater than in the case of sinusoidal signal reconstruction. The Fractional Fourier transform can be efficiently used for chirps, but not for the signals with nonlinear frequency laws (e.g., a cubic phase function that appears in radars).

In this paper, we will first consider the Polynomial Fourier transform (PFT) [37], which produces a sparse representation for any polynomial phase signals. The proposed theory is further extended to nonstationary signals using the Local polynomial Fourier transform (LPFT) in signal reconstruction [38]. It is shown that by adopting and applying CS algorithms in the PFT or LPFT domain, signals with polynomial phases can be accurately reconstructed. The presented theory is demonstrated on various numerical examples. Finally, we provide the analysis of the reconstruction accuracy in noisy environment in terms of the number of available observations.

The paper is organized as follows. The theoretical background regarding the PFT is presented in Section II. A description of the proposed compressive sensing approach is given in Section III. The experimental verification of the proposed approach is given through numerical examples in Section IV.

This work is supported by the Montenegrin Ministry of Science, project grant CS-ICT, "New ICT Compressive sensing based trends applied to: multimedia, biomedicine and communications".

The authors are with the Faculty of Electrical Engineering, University of Montenegro, Dzordza Vasingtona bb, 81000 Podgorica, Montenegro (e-mail: srdjan@ac.me, irenao@ac.me, ljubisa@ac.me)

## II. Compressive Sensing in the Polynomial Fourier Domain

Compressive sensing algorithms are used for signals with sparse representations in a certain domain. The Fourier analysis reveals sinusoidal behavior, so it can be used for the reconstruction of signals with linear phase. However, when dealing with nonlinear polynomial phase signals, a sparse signal representation can be achieved by using the PFT. Namely, the PFT can be used to demodulate components of interest, reducing them to the sparse sinusoidal components.

For instance, consider a polynomial phase signal $x(t) = \sum_{i=1}^{K} r_i e^{j\Phi_i(t)}$, where $r_i$ is the amplitude and $\Phi_i(t) = a_{1i}t + ... + a_{ni}t^n / n!$ is the phase of the $i$-th component. In general, the PFT of $x(t)$ can be defined as follows:

$$X(\omega_1,...,\omega_n) = \sum_{i=1}^{K} r_i \int_{-\infty}^{\infty} e^{j((a_{1i}-\omega_1)t+ ... + \frac{(a_{ni}-\omega_n)}{n!}t^n)} dt. \quad (1)$$

For all points in the n-dimensional frequency space that correspond to the positions of signal components characterized by: $\omega_1 = a_{1i}$, ..., $\omega_n = a_{ni}$, $i=1,...,K$ we obtain:

$$X(\omega_1,...,\omega_n) = \sum_{i=1}^{K} 2\pi r_i \delta(\omega_1 - a_{1i}, ..., \omega_n - a_{1n}) \quad (2)$$

In other words, $X(\omega_1,...,\omega_n) \to \infty$ at the positions of signal components: $\omega_1 = a_{1i}$, ..., $\omega_n = a_{ni}$. Thus, ideally, $X$ can be considered as a sparse $K$-component representation. Otherwise, for $\omega_1 \neq a_{1i}$, ..., $\omega_n \neq a_{ni}$, $i=1,...,K$, (1) has finite value, negligible when compared to (2).

### A. Signal sparsity in the discrete PFT domain

Let us observe the signal vector **x**, with elements $x(m)$, consisting of a sum of $K$ polynomial phase signals:

$$x(m) = \sum_{i=1}^{K} x_i(m) = \sum_{i=1}^{K} r_i e^{j\frac{2\pi}{M}\left(ma_{1i} + \frac{m^2 a_{2i}}{2} + ... + \frac{m^n a_{ni}}{n!}\right)}, \quad (3)$$

where the polynomial coefficients are assumed to be bounded integers. In order to provide a sparse signal representation, we will use the discrete form of the PFT, which can be defined as:

$$X(k_1,...,k_n) = \sum_{m=0}^{M-1} \sum_{i=1}^{K} r_i e^{j\frac{2\pi}{M}\left(ma_{1i} + \frac{m^2 a_{2i}}{2} + ... + \frac{m^n a_{ni}}{n!}\right)} \\ \times e^{-j\frac{2\pi}{M}(\frac{m^2 k_2}{2} + ... + \frac{m^n k_n}{n!})} e^{-j\frac{2\pi}{M}mk_1}. \quad (4)$$

When the set of PFT parameters $(k_2, k_3, ..., k_n)$ is chosen to match the polynomial phase coefficients $(a_{2i}, a_{3i}, ..., a_{ni})$:

$$(k_2, k_3, ..., k_n) = (a_{2i}, a_{3i}, ..., a_{ni}), \quad (5)$$

the $i$-th signal component is demodulated and the sinusoid: $r_i e^{j2\pi/M(ma_{1i})}$ becomes dominant in the PFT spectrum. In this case, the spectrum is highly concentrated at $k = a_{1i}$. Otherwise, when $(k_2, k_3, ..., k_n) \neq (a_{2i}, a_{3i}, ..., a_{ni})$ for all $i \in (1,...,K)$, the spectrum is dispersed. Here, we need to make the assumption: $M|r_{\min}| > \sum_{i=1}^{K} |r_i|$, where $r_{\min}$ is the minimum amplitude of the components. Consequently, we search for $K$ sets of parameters $\{(k_{2i}, k_{3i}, ..., k_{ni}) | i=1,...,K\}$ such that $(k_{2i}, k_{3i}, ..., k_{ni}) = (a_{2i}, a_{3i}, ..., a_{ni})$ holds, and $X(k_{1i},...,k_{ni})$ is highly concentrated. In other words, these parameters are chosen to provide the best concentrated vector $X(k_1, k_2, ..., k_n)$ for all $i=1,...,K$. The maximum component will be located at the frequency $k = a_{1i}$. Finally, in real applications, the polynomial phase signals could be considered as compressible rather than strictly sparse in the PFT domain.

## III. Reconstruction of CS signals in the PFT domain

In the previous analysis it was shown that a polynomial phase signal exhibits certain sparsity for a properly chosen set of the PFT parameters. This makes it convenient for the application of CS reconstruction methods. However, it is necessary to define a specific reconstruction procedure which will include also the parameters search and the signal demodulation approach.

In order to be compliant with the CS notations, let us rewrite the PFT in the matrix form using the notations:

$$\mathbf{X}_s = \mathcal{F}_M \mathbf{s}, \quad (6)$$

where $\mathbf{X}_s = [X(0), X(1),...,X(M-1)]^T$ is the vector of Fourier transform coefficients of the signal $\mathbf{s} = [s(0), s(1),...,s(M-1)]^T$, which contains samples $x(m)$ multiplied by exponential terms from the vector $\boldsymbol{\varphi} = [\varphi(0), \varphi(1),...,\varphi(M-1)]^T$ (the component-wise multiplication is denoted by ($\circ$)):

$$\mathbf{s} = \mathbf{x} \circ \boldsymbol{\varphi}, \quad (7)$$

where the components of $\boldsymbol{\varphi}$ reads as follows,

$$\varphi(m) = \exp(-j2\pi/M(k_2 m^2/2 + ... + k_n m^n/n!)), \ m \in (0,..,M-1)$$

The discrete Fourier transform matrix of size $M \times M$ is denoted by $\mathcal{F}_M$, with elements:

$$\mathcal{F}(m,k) = e^{-j2\pi km/M},$$
$$m = 0,...,M-1, \ k = 0,...,M-1 \quad (8)$$

For a chosen set of parameters $(k_2, k_3, ..., k_n)$ in $\boldsymbol{\varphi}$ that is equal to the set $(a_{2i}, a_{3i}, ..., a_{ni})$ in **x**, $\mathbf{X}_s = \mathbf{X}_{si}$ is characterized by one dominant sinusoidal component at the frequency $a_{1i}$. Now, assume that **x** is compressively sampled and represented by only $N$ random measurements. Thus, instead of **x** we are dealing with a measurement vector **y** obtained using the incoherent measurement matrix **Φ** of size $N\times M$:

$$\mathbf{y}=\mathbf{\Phi s}=\mathbf{\Phi}(\mathbf{x}\circ\mathbf{\varphi}). \qquad (9)$$

Now, using (6) the previous equation can be written as:

$$\mathbf{y}=\mathbf{\Phi}\mathcal{F}_M^{-1}\mathbf{X_s}=\mathbf{A}_{cs}\mathbf{X}_s, \qquad (10)$$

where $\mathbf{A}_{cs}=\mathbf{\Phi}\mathcal{F}_M^{-1}$. Note that $\mathcal{F}_M$ is the standard Fourier transform matrix, while **Φ** is a random matrix such that $\mathbf{A}_{cs}$ results in a random sets of rows from a discrete Fourier transform matrix, as in [39]. It has been known that such random partial Fourier matrix lead to fast recovery algorithms, and it satisfies a near optimal RIP with high probability [40]-[42].

When $(k_2, k_3, ..., k_n)=(a_{2i}, a_{3i}, ..., a_{ni})$ then $\mathbf{X}_s$ can be observed as a demodulated version of the *i*-th signal component $\mathbf{X}_s=\mathbf{X}_{si}$ that can be recovered by solving the $l_1$-norm minimization problem in the form:

$$min\|\mathbf{X}_{si}\|_1 \text{ subject to } \mathbf{y}=\mathbf{A}_{cs}\mathbf{X}_{si}. \qquad (11)$$

However, in the case of multicomponent signals the inter-components products appear and act as spectral noise that may complicate the reconstruction of the *i*-th signal component. In that case, a sparse solution of (11) can be provided using the threshold based single iteration algorithm [45]. The threshold is derived to detect sparse signal components in the non-sparse noisy spectrum caused by missing samples. The same approach can be applied here as well. Namely, the algorithm will easily detect and reconstruct the dominant *i*-th component $\mathbf{X}_s=\mathbf{X}_{si}$ with the amplitude $r_i$, while leaving other insignificant components below the threshold. Also, when $(k_2,...,k_n)\neq(a_{2i},...,a_{ni})$ and $N\ll M$ there is no a dominant sinusoidal component and the algorithm returns zero value. Consequently, we can obtain the *i*-the component of a signal **s** as the inverse Fourier transform of $\mathbf{X}_{si}$, while the original *i*-th component of **x** results from re-modulation: $\mathbf{x}=\mathbf{\varphi}^{-1}\mathbf{s}$, where $\mathbf{\varphi}^{-1}$ is a componentwise inverse of $\mathbf{\varphi}$. The procedure can be summarized by the flowchart given in Fig. 1.

As we change the values of parameters $k_2,...,k_n$ between $k_{min}$ and $k_{max}$, one by one component will be detected through iterations. Hence, the number of components does not have to be known in advance.

The reconstructed components amplitudes at frequencies $k_{1i}$, for $i=1,...,K$, could be close to the exact values $r_1, r_2, ..., r_K$, but still the amplitude correction needs to be performed. Let us denote the set of measurements positions (selected by **Φ**) as $(m_1, m_2,..., m_N)$, while the detected signal phase parameters are: $\mathbf{k_i}=(k_{1i}, k_{2i}, ..., k_{ni})=(a_{1i}, a_{2i}, ..., a_{ni})$. In order to calculate the exact amplitudes of signal components, we observe the set of equations in the form:

$$\begin{bmatrix} x(m_1) \\ ... \\ ... \\ x(m_N) \end{bmatrix} = \begin{bmatrix} e^{j\frac{2\pi}{M}(m_1 k_{11}+...+m_1^n k_{n1})} & ... & e^{j\frac{2\pi}{M}(m_1 k_{1K}+...+m_1^n k_{nK})} \\ e^{j\frac{2\pi}{M}(m_2 k_{11}+...+m_2^n k_{n1})} & ... & e^{j\frac{2\pi}{M}(m_2 k_{1K}+...+m_2^n k_{nK})} \\ ... & ... & ... \\ e^{j\frac{2\pi}{M}(m_N k_{11}+...+m_N^n k_{n1})} & ... & e^{j\frac{2\pi}{M}(m_N k_{1K}+...+m_N^n k_{nK})} \end{bmatrix} \begin{bmatrix} r_1 \\ ... \\ ... \\ r_K \end{bmatrix}$$
(12)

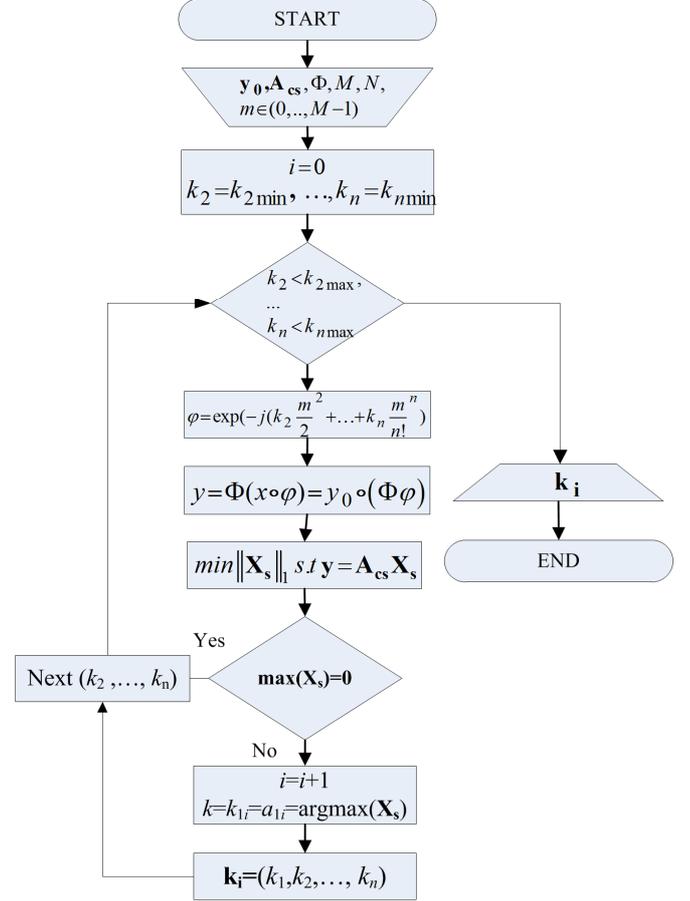

**Fig. 1 Parameters estimation procedure for CS reconstruction used in this paper**

or in other words: **y=AX**. The vector of measurements is $\mathbf{y}=[x(m_1), ..., x(m_N)]^T$ and the vector $\mathbf{X}=[r_1,...,r_K]^T$ contains the desired $K$ signal amplitudes. The matrix **A** represents a new CS matrix of size $N\times K$ which is based on the PFT:

$$\mathbf{A}= \begin{bmatrix} e^{j\frac{2\pi}{M}(m_1 k_{11}+...+m_1^n k_{n1})} & ... & e^{j\frac{2\pi}{M}(m_1 k_{1K}+...+m_1^n k_{nK})} \\ e^{j\frac{2\pi}{M}(m_2 k_{11}+...+m_2^n k_{n1})} & ... & e^{j\frac{2\pi}{M}(m_2 k_{1K}+...+m_2^n k_{nK})} \\ ... & ... & ... \\ e^{j\frac{2\pi}{M}(m_N k_{11}+...+m_N^n k_{n1})} & ... & e^{j\frac{2\pi}{M}(m_N k_{1K}+...+m_N^n k_{nK})} \end{bmatrix}$$
(13)

The rows of **A** correspond to measurements $(m_1,m_2,...,m_N)$, and columns correspond to frequencies $\mathbf{k_i}=(k_{1i}, k_{2i},..., k_{ni})=(a_{1i}, a_{2i}, ..., a_{ni})$, for $i=1,...,K$. The solution of the observed problem can be obtained as:

$$\mathbf{X}=(\mathbf{A}^H\mathbf{A})^{-1}\mathbf{A}^H\mathbf{y}$$

The resulting reconstructed signal is obtained as:

$$x(m)=r_1 e^{-j\frac{2\pi}{M}\left(ma_{11}+\ldots+\frac{m^n a_{n1}}{n!}\right)}+\ldots+r_K e^{-j\frac{2\pi}{M}\left(ma_{1K}+\ldots+\frac{m^n a_{nK}}{n!}\right)}.$$

### A. Analyzing the accuracy in the noisy signal case

Now, let us consider a case when the available samples are corrupted by additive noise $\varepsilon(n)$. When the recovery is achieved, the accuracy is related to the input additive noise and the number of available samples. The measurement vector in the case of noisy observation can be written as:

$$\mathbf{y}+\boldsymbol{\varepsilon}=\mathbf{AX}. \qquad (14)$$

Then the solution can be obtained as follows:

$$\mathbf{X}=\left(\mathbf{A}^H\mathbf{A}\right)^{-1}\mathbf{A}^H(\mathbf{y}+\boldsymbol{\varepsilon}), \qquad (15)$$

where the operator $(^H)$ denotes the conjugate transpose. Having in mind that the Fourier transform vector $\mathbf{X}$ consists of the signal and noise parts: $\mathbf{X}=\mathbf{X_s}+\mathbf{X_n}$, we can write:

$$\mathbf{X}_s=\left(\mathbf{A}^H\mathbf{A}\right)^{-1}\mathbf{A}^H\mathbf{y}, \quad \mathbf{X}_n=\left(\mathbf{A}^H\mathbf{A}\right)^{-1}\mathbf{A}^H\boldsymbol{\varepsilon}. \qquad (16)$$

After recovering $\mathbf{X}_n$, the full set of noise samples can be obtained as the inverse Fourier transform of $\mathbf{X}_n$. The reconstructed transform vector of the observed noise is obtained using only $N$ samples (observations), and thus should be scaled by $M/N$ in order to correspond to the full set of $M$ samples. Consequently, the signal to noise ratio (SNR) is:

$$SNR_{in}=10\log\frac{\sum_{n=0}^{M-1}|x(n)|^2}{\frac{K}{M}\left(\frac{M^2}{N^2}\right)\sum_{n=m_1}^{m_N}|\varepsilon(n)|^2}=10\log\frac{\sum_{n=0}^{M-1}|x(n)|^2}{\frac{K}{N}\sum_{n=0}^{M-1}|\varepsilon(n)|^2} \qquad (17)$$

where $\frac{1}{N}\sum_{n=m_1}^{m_{Na}}|\varepsilon(n)|^2=\frac{1}{M}\sum_{n=0}^{M-1}|\varepsilon(n)|^2$ holds, while the energy of $\varepsilon(n)$ is additionally scaled by $K/M$ since only $K$ ($K$ is the number of frequency components) out of $M$ coefficients are used in the reconstruction. Finally, the resulting SNR can be obtained as:

$$SNR_{out}=10\log\frac{\sum_{n=0}^{M-1}|x(n)|^2}{\sum_{n=0}^{M-1}|\varepsilon(n)|^2}-10\log\frac{K}{N}=SNR_{in}-10\log\frac{K}{N}. \quad (18)$$

The resulting SNR is a function of the sparsity level and the number of available samples.

### B. CS using the Local polynomial Fourier transform

In the case of signals with non-stationary phase functions, the proposed approach can be adapted and extended using the LPFT, which represents the windowed PFT. Namely, at a time instant $m$, the LPFT is calculated as the Fourier transform of windowed signal: $Z_M(m)=FT\{x_w(m)\}=FT\{x(m)w(m-n)\}$, resulting in a vector of frequency coefficients. For the simplicity, we consider the non-overlapping windows. After calculating LPFT vectors for all time instants, we can write a single matrix equation as follows:

$$\mathbf{Z}=\mathcal{F}_{M,N}\mathbf{s}_w \qquad (19)$$

such that the LPFT vector $\mathbf{Z}$ is composed of vectors: $\mathbf{Z}_M(0)$, $\mathbf{Z}_M(M)$, …, $\mathbf{Z}_M(N-M)$. The matrix $\mathcal{F}_{M,N}$ of size $N\times N$ is obtained as a Kronecker product:

$$\mathcal{F}_{M,N}=\mathbf{I}_{N/M}\otimes\mathcal{F}_M,$$

where $\mathbf{I}_{N/M}$ denotes the identity matrix of size $(N/M)\times(N/M)$, and $N$ is a multiple of $M$.

Instead of the full windowed demodulated signal $\mathbf{s}_w$, we deal with an incomplete set of samples or observations in the form: $\mathbf{y}=\boldsymbol{\Phi}\mathbf{s}_w$. Then, $\mathbf{s}_w$ can be written using $\mathbf{Z}$:

$$\mathbf{y}=\boldsymbol{\Phi}\mathcal{F}_{M,N}^{-1}\mathbf{Z} \qquad (20)$$

where the CS matrix is given by: $\mathbf{A}_{cs}=\boldsymbol{\Phi}\mathcal{F}_{M,N}^{-1}$.

In order to detect set $(k_2,k_3,\ldots,k_n)$ leading to the best concentrated LPFT spectrum, the CS minimization problem can be observed as follows:

$$min\|\mathbf{Z}\|_1 \text{ subject to } \mathbf{y}=\mathbf{A}_{cs}\mathbf{Z}. \qquad (21)$$

For a properly chosen set $(k_2,k_3,\ldots,k_n)$ corresponding to one of the signal components, the algorithm [45] is used to detect dominant component in $\mathbf{Z}_M(p)$ and its frequency $k=k_1$. After detecting all signal components, the reconstructed version of $\mathbf{Z}$ is obtained as:

$$\mathbf{Z}=(\widetilde{\mathbf{A}}\widetilde{\mathbf{A}}^H)^{-1}\widetilde{\mathbf{A}}^H\mathbf{y}, \qquad (22)$$

where $\widetilde{\mathbf{A}}$ is a new CS matrix obtained from: $\left(\mathbf{I}_{N/M}\otimes\mathbf{A}\right)$, by taking the rows corresponding to available measurements, and columns corresponding to the frequencies $\mathbf{k_i}=(k_{1i}, k_{2i}, \ldots, k_{ni})$, for $i=1,\ldots,K$. The matrix $\mathbf{A}$ is of size $M\times M$ obtained as in (13).

## IV. NUMERICAL EXAMPLES

***Example 1:*** Let us consider a third order polynomial phase signal in the form:

$$x(t)=\exp(-j2\pi 16Tt^3+j8\pi Tt)$$

where $t=[-1/2,1/2)$ with step $\Delta t=1/1024$, while $T=32$. Thus, the total signal length is $M=1024$ samples. Its discrete-time version is:

$$x(m)=\exp(j\frac{2\pi}{1024}(-m^3\frac{1}{2048}+128m))$$

with $-512\leq m<512$.

The signal is represented by $N=32$ randomly chosen measurements (3% of the total signal length). The incomplete set of samples is first multiplied by the corresponding exponential term $\exp(j2\pi\alpha Tt^3)$, in order to perform demodulation. Namely, different values of parameters $\alpha$ are used (between $\alpha_{min}=-20T=-640$ to $\alpha_{max}=20T=640$). After multiplication, the available samples are used for signal reconstruction (single iteration procedure [45], [46] is applied). It is important to emphasize that the resulting signal (multiplied by exponential terms of third order) will be spread in the PFT domain, as long as $\alpha$ is not matched by the signal parameter. Consequently, the single iteration reconstruction procedure will not return any relevant component (Fig.2 a and b).

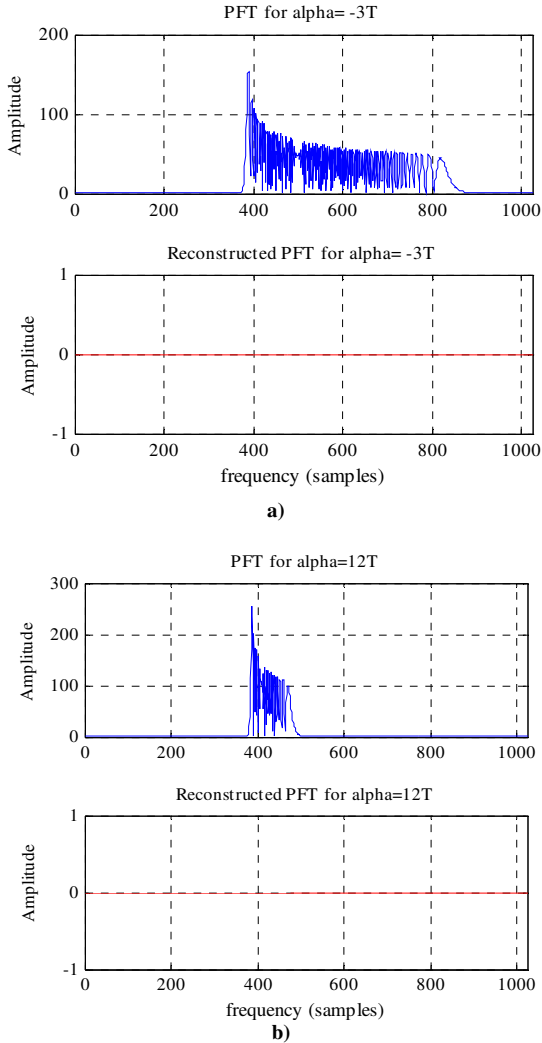

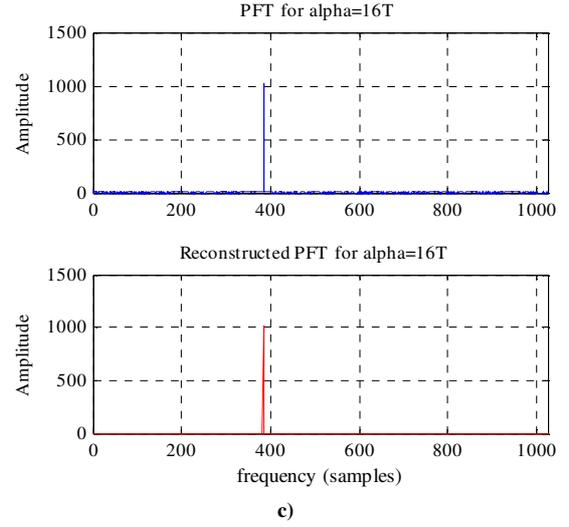

**Fig 2.** The PFT (full data set) and reconstructed PFT for different values of $\alpha$: a) $\alpha=-3T$, b) $\alpha=12T$, c) the result for $\alpha=16T$ ($\alpha$ is matched with signal parameter value)

The signal becomes highly concentrated and sparse when $\alpha$ is equal to the chirp rate, allowing successful reconstruction (Fig.2.c). The reconstructed PFTs are calculated for each $\alpha$ in the range $\alpha_{min}$ to $\alpha_{max}$. The third order phase parameter is obtained by changing value of parameter $\alpha$ and calculating reconstructed PFT. The position of maximum determines the exact parameter value. In this example the position of maximum is 37 (Fig. 3) which corresponds to $\alpha=16T$, while for other $\alpha$ between $-20T$ and $20T$, the reconstruction results are zero-values.

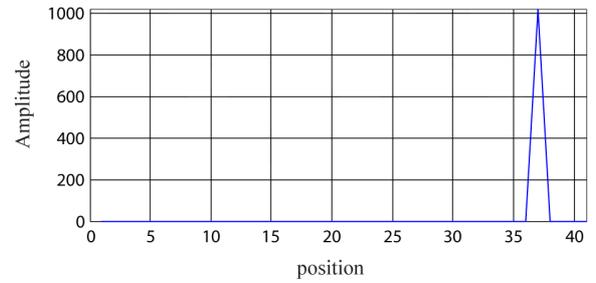

**Fig 3.** The maximal values of PFT obtained for 41 different value of $\alpha$ ($\alpha_{min}=-20T$ to $\alpha_{max}=20T$)

*Example 2:* In this example we observe a multicomponent signal that consists of two chirps:

$$x(t)=\exp(-j2\pi 8Tt^2+j8\pi Tt)+\exp(-j2\pi 16Tt^2).$$

The signal parameters $t$, $\Delta t$ and $T$, as well as the number of available samples are the same as in the previous example. In order to perform demodulation, the incomplete set of samples is firstly multiplied by the corresponding exponential term $\exp(j2\pi\alpha Tt^2)$. Observe that in this case we deal with two different chirp rates $8T$ and $16T$ ($T=32$), which cannot be detected at once. Hence, we need to change values of $\alpha$ ($\alpha_{min}=-20T$ to $\alpha_{max}=20T$, with the step $T$), aiming to detect each chirp rate. Note that in practical applications the domain of chirp-rates can be determined according to a priori knowledge of the application set up (for example in radars). Let us first observe the spectral representations obtained using the PFT for different values of $\alpha$ (Fig. 4). For the purpose of analysis, the illustrations are made for the full set of signal samples. We might see from Fig. 3, that for $\alpha$ equal to one of the components chirp rate, we obtain very concentrated representation (almost a sinusoid). Now, we need to provide an efficient procedure for detection of chirp rates.

the signal is represented by a small set of samples. The LPFT is calculated using non-overlapping windows of size 32 samples, and for different $\alpha$ values. Thus, for each window, the available signal samples are multiplied with the corresponding exponential, and the signal is reconstruction is performed, resulting in the LPFT. The procedure is repeated for each windowed signal part. The LPFT obtained for different $\alpha$ are shown in Fig. 6. Note that for the cases $\alpha=8T$ and $\alpha=14T$ we obtain two pure sinusoids in the LPFT domain, meaning that $\alpha$ is matched with the chirp rate.

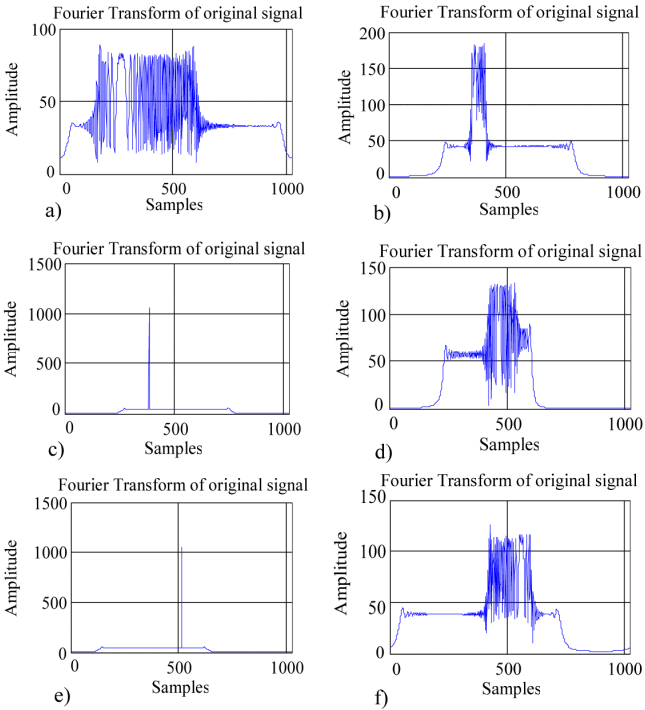

Fig. 4. The spectral representations obtained for different values of $\alpha$. The cases: c) $\alpha=8T$ and e) $\alpha=16T$, illustrate the results when the chirp rates are matched (concentrated representations)

In each iteration the single iteration reconstruction procedure is applied, [45], [46]. When the PFT spectrum is spread ($\alpha$ does not match the chirp rate), the result of reconstruction is zero, or it is much smaller than in the case of concentrated spectrum ($\alpha$ matches the chirp rate). The two highest components obtained for the range $\alpha \in [\alpha_{min}, \alpha_{max}]$ correspond to the cases when desired chirp rates are matched. After detecting the chirps rates, the original amplitudes of demodulated components are firstly recovered and then the chirp components are obtained using the re-modulation process. The two recovered signal components are shown in Fig. 5, as well as the error between the original and reconstructed component, which is completely negligible.

*Example 3:* Consider a multicomponent signal:

$$x(t)=\begin{cases} exp(-j2\pi T \cdot 8t^2 + j8\pi Tt), & t\in[-\frac{1}{2},0] \\ exp(-j2\pi T \cdot 14t^2), & t\in(0,\frac{1}{2}] \end{cases}$$

where the total number of samples is $M=1024$ ($t=[-1/2,1/2]$, with the step $\Delta t=1/1024$). Due to the compressive sampling,

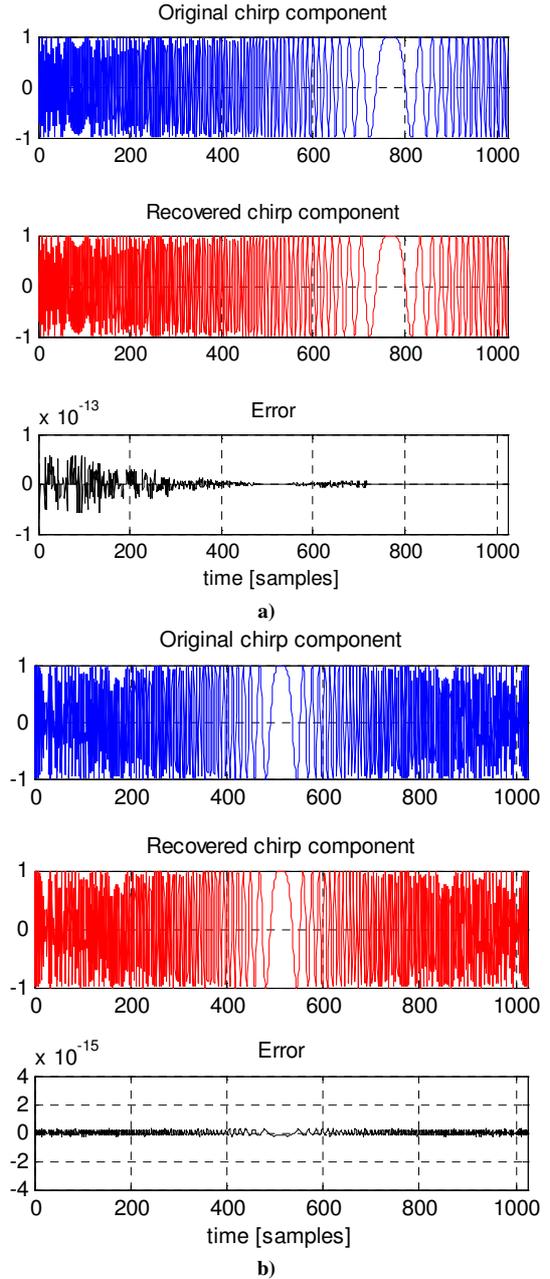

Fig. 5. a) First chirp component: real parts of original, reconstructed component and reconstruction error, b) Second chirp component: real parts of original, reconstructed component and reconstruction error

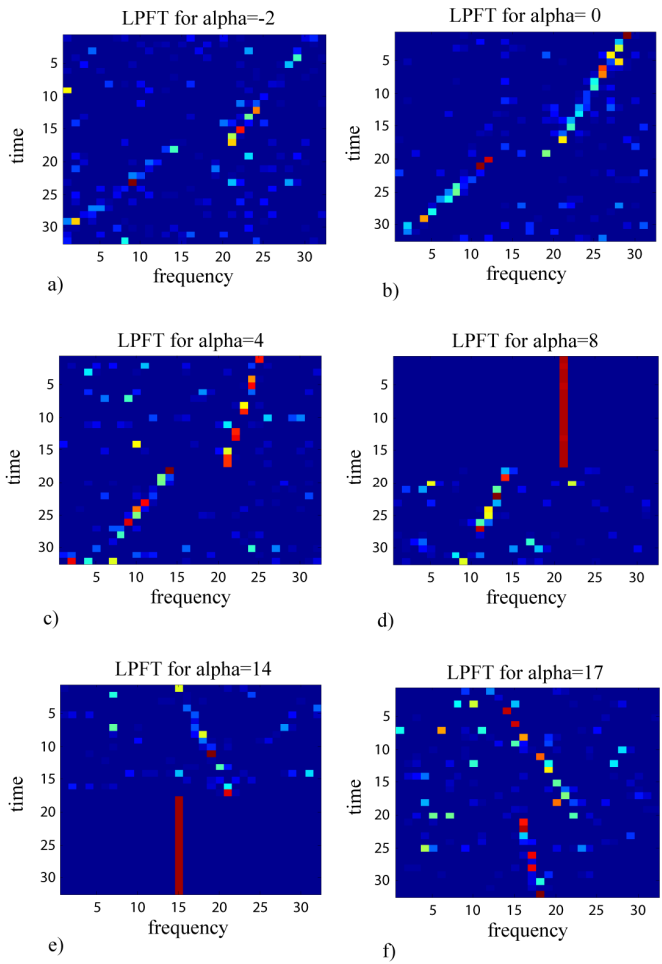

**Fig 6. LPFT for different α, where cases d) and e) correspond to α=8T and α=14T. Number of available samples is 25% of the window length**

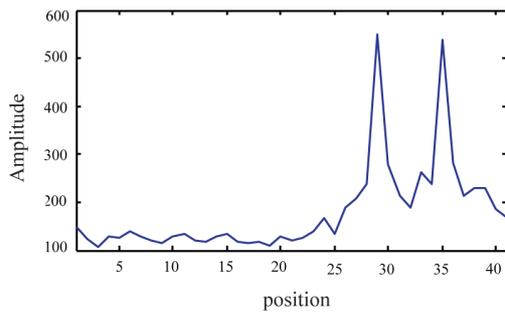

**Fig 7. Maximal components obtained from reconstruction algorithm for different values of α ($α_{min}$=-20T to $α_{max}$=20T)**

For each value of α we firstly calculate the projection of LPFT onto the frequency axes and then collect the maximal component of the projection vector. The two highest peaks in different iteration will correspond to the matched chirp rates cases, as shown in Fig. 7 (position 29 corresponds to α=8T, while position 35 corresponds to α=14T).

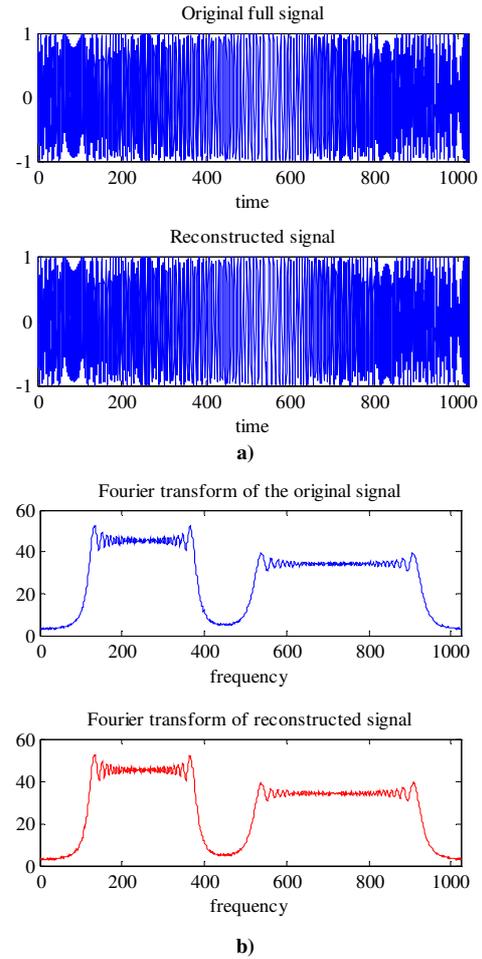

**Fig. 8. a) Original (full data set) and reconstructed signals in time domain- real part, b) The Fourier transform of the original and reconstructed signals – absolute values**

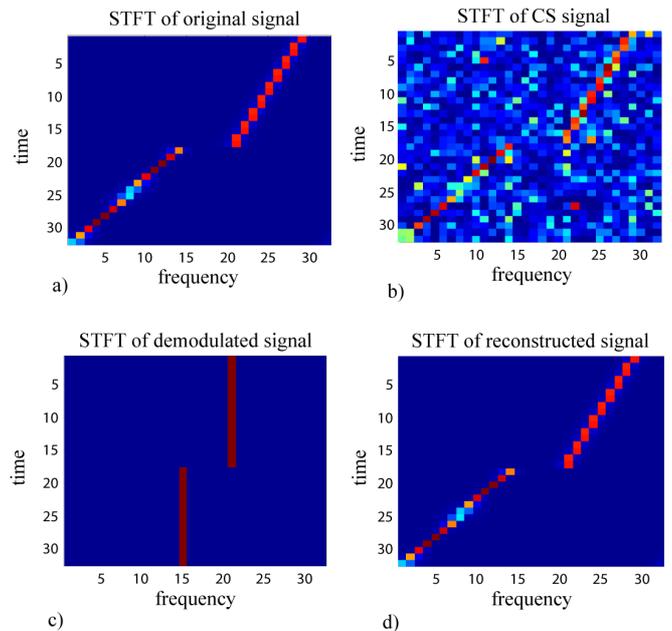

**Fig. 9. a) STFT of the original full signal, b) STFT of the CS signal (25% of samples are available), c) STFT of the demodulated signal, d) STFT of the reconstructed signal**

The original (full data) and reconstructed signals along with the corresponding Fourier transforms are shown in Fig. 8. The STFTs of the original full signal, the CS signal, the demodulated signal, and the reconstructed signal are shown in Fig. 9.

***Example 4:*** In this example we will observe the noisy signal case, in order to measure the output SNR ($SNR_{out}$). Hence, we observe the signal with $K=3$ components in the form:

$$x(t)=\exp(-j2\pi 8Tt^2+j8\pi Tt)+\exp(-j2\pi 8Tt^2+j16\pi Tt)+$$
$$+\exp(-j2\pi 8Tt^2-j8\pi Tt)+\varepsilon(t)$$

where $\varepsilon(t)$ is Gaussian noise. The same signal parameters are used in this example as well. The statistical evaluation on the achieved $SNR_{out}$ is done within 1000 repetitions of the reconstruction procedure (for a predefined $SNR_{in}\approx 5$dB and $SNR_{in}\approx 10$ dB). The theoretical and statistical (experimental) results are given in Table 1 (columns III and IV, respectively). It can be observed that for each pair ($K,N$), the $SNR_{out}$ is reduced for $10\log(K/N)$ compared to $SNR_{in}$.

TABLE 1. THE EXPERIMENTAL RESULTS FOR THE $SNR_{OUT}$

| No. measur. | $SNR_{in}$ | $SNR_{out}=SNR_{in}-10\log(K/N)$ | $SNR_{out}$ statistically |
|---|---|---|---|
| $N=256$ | $SNR_{in}=$ 5 dB | 24 dB | 24.3 dB |
| $N=80$ |  | 19.3 dB | 19.62 dB |
| $N=256$ | $SNR_{in}=$ 10 dB | 29.3 dB | 29.64 dB |
| $N=80$ |  | 24.25 dB | 24.04 dB |

***Example 5:*** The performance of the proposed approach is finally examined via statistical test method to illustrate the convergence of the algorithm. The statistical test included several realizations with: a) a random selection of phase parameters, b) a different number of signal components $K$ (up to 64), and c) a different number of available samples ($N$). Some interesting results are shown in Fig.10 and Fig.11 (for $n=2$; $M=128$ and $M=512$, respectively). Based on these statistical tests, we may conclude that the algorithm in this case converges to the exact solution for approximately $N\geq 6K$ with a high probability.

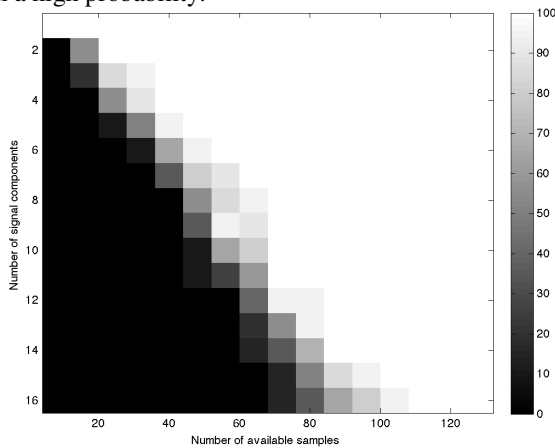

**Fig. 10 The percentage of the full reconstruction (with error bellow $10^{-10}$) in terms of the number of measurements and the number of signal components (full signal length $M=128$)**

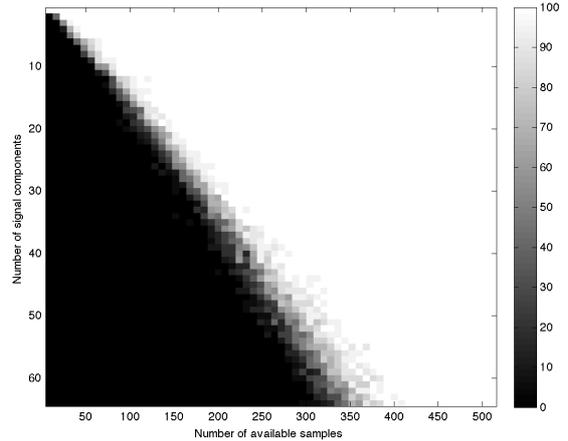

**Fig. 11 The percentage of the full reconstruction (with error bellow $10^{-10}$) in terms of the number of measurements and the number of signal components (full signal length $M=512$)**

## V. CONCLUSION

The method for compressive sensing reconstruction of nonstationary signals with polynomial phase function is proposed. The PFT and the LPFT are used to achieve the sparse representations of considered signals. It was shown that the polynomial phase signals could be stationarized and sparse in these domains, which further allows reconstruction from random measurements. Moreover, the new PFT based CS matrix was derived, as well as the signal reconstruction algorithm that includes phase parameters detection, demodulation and components reconstruction. Note that the exact signal reconstruction (reconstructed signal is identical to the original one) can be achieved in the non-noisy signal case. In noisy environment, the SNR at the output is decreased for $10\log K/N$ comparing to the input SNR. The experimental results showed that the proposed method can be efficiently applied in different scenarios: highly undersampled data (3% of the original signal length), a noisy signal, and a signal that changes phase parameters over its duration. The future work may include signals characterized by other phase laws, which may further lead to the proposed concept generalization.

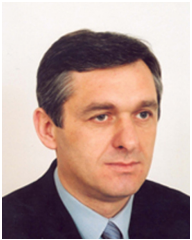

**Srdjan Stanković** (M'94-SM'08) received M.S. degree from the University of Zagreb, Croatia, in 1991, and the Ph.D. degree from the University of Montenegro (UoM) in 1993, both in Electrical Engineering (EE). In 1992 he joined the Faculty of EE, UoM, where he is currently a Full Professor. In the period 2007-2013 he served as Dean of this Faculty. His interests are in signal processing and multimedia systems. He is a member of the Board of Directors in Montenegrin Broadcasting Company since 2004, where he was also the President (2005- 2006). In 1998 he as at the Department of Informatics, Aristotle University, Thessaloniki. In the 1999-2000, he was on leave at the Darmstadt University of Technology, with the Signal Theory Group, and in 2002 he spent three months at the Department of Computer Science, University of Applied Sciences Bonn-Rhein-Sieg, supported by the AvH Foundation. From 2004 to 2010, he had research stays with the E3I2 Lab, ENSIETA, Brest, France, Center for DSP research at King's College London, Laboratory of Mathematical Methods in Image Processing, at Moscow State Lomonosov University, GIPSA Laboratory at INPG Grenoble. He spent academic 2012/2013 with the Center for Advanced Communications at the Villanova University, PA. He published a book "Multimedia Signals and Systems" by Springer and several textbooks on electronics devices (in Montenegrin). He published more than 200 journal and conference papers. In 2010, he was the Lead Guest Editor of the EURASIP Journal on Advances in Signal Processing for the special issue: Time-frequency analysis and its applications to multimedia signals, as well as a Guest Editor of the Signal Processing for special issue: Fourier related transforms. He is the Lead Guest Editor of the IET Signal Processing for the Special issue: Compressive Sensing and Robust Transforms. From 2005 to 2009 he was serving as an Associate Editor of the IEEE Transactions on Image Processing. In 2011 he was awarded by the Ministry of Science in Montenegro as the Leader of the Best Scientific Project in Montenegro.

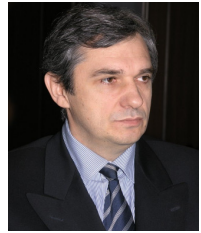

**Ljubiša Stanković** (*M'91–SM'96–F'12*) was born in Montenegro in 1960. He received the B.S. degree in EE from the University of Montenegro (UoM), the M.S. degree in Communications from the University of Belgrade and the Ph.D. in Theory of Electromagnetic Waves from the UoM. As a Fulbright grantee, he spent 1984-1985 academic year at the Worcester Polytechnic Institute, USA. Since 1982, he has been on the faculty at the UoM, where he has been a full professor since 1995. In 1997-1999, he was on leave at the Ruhr University Bochum, Germany, supported by the AvH Foundation. At the beginning of 2001, he was at the Technische Universiteit Eindhoven, The Netherlands, as a visiting professor. He was vicepresident of Montenegro 1989-90. During the period of 2003-2008, he was Rector of the UoM. He is Ambassador of Montenegro to the UK, Ireland and Iceland. His current interests are in Signal Processing. He published about 350 technical papers, more than 120 of them in the leading journals, mainly the IEEE editions. Prof. Stanković received the highest state award of Montenegro in 1997, for scientific achievements. He was a member the IEEE SPS Technical Committee on Theory and Methods, an Associate Editor of the *IEEE Transactions on Image Processing*, the *IEEE Signal Processing Letters* and numerous special issues of journals. Prof. Stanković is an Associate Editor of the *IEEE Transactions on Signal Processing*. He is a member of the National Academy of Science and Arts of Montenegro (CANU) since 1996 and a member of the European Academy of Sciences and Arts.

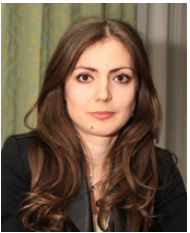

**Irena Orović** was born in Montenegro, in 1983. She received the B.Sc., M.Sc., and Ph.D. degrees in electrical engineering from the University of Montenegro (UoM), in 2005, 2006, and 2010, respectively. From 2005 to 2010, she was a TA with the UoM. Since 2010, she has been an Assistant Professor with the Faculty of EE, UoM. She finished her Diploma project in ENSIETA Brest, France. She received the Award of the city Podgorica as the best student at the Faculty of EE, UoM in 2003, the Award of Montenegrin Academy of Science and Arts (CANU) in 2004, the Award for the best student of natural and technical sciences at the UoM 2005, TRIMO Awards Slovenia - Award for the best PhD thesis in 2010, Award for the Best Woman Scientist in Montenegro, Ministry of Science of Montenegro, 2011. She has spent a period of time in ENSIETA Bresta, France during 2005 and 2006. In 2007 she spent one month at the University Bonn-Rhien Sieg in Bonn, Germany. During 2008 and 2009 she stayed several times at INPG Grenoble, France (2008 and 2009), and during 2010 and 2011 within the Villanova Univerzitet USA. Dr Orovic has published more than 60 papers in the leading scientific journals and conferences. As a co-author she has published 6 books. She is Vice President of the Council for Scientific Research Activity on Montenegro. Her research interests include compressive sensing, multimedia signals and systems, and time-frequency analysis with applications.